\documentclass[article,twocolumn]{revtex4}
\usepackage{graphicx}
\usepackage{amssymb}
\usepackage{bm}
\usepackage{hyperref}%for .pdf
\usepackage{epstopdf}%for .pdf

\begin{document}\large

\title{Influence from cosmological uncertainties on galaxy number count at faint limit}

\author{Keji Shen$^{1}$}
\author{Qiang Zhang$^{1}$}
\author{Xin-he Meng$^{1,2}$}
\email{xhm@nankai.edu.cn}

\affiliation{
$^1$Department of Physics, Nankai University, Tianjin 300071, China.\\
}
\affiliation{
$^2$State Key Laboratory of Theoretical Physics, Institute of Theoretical Physics, Chinese Academic of Science, Beijing 100190, China.\\
}

\date{\today}

\begin{abstract}
Counting galaxy number density with wide range sky surveys has been well adopted in researches focusing on revealing evolution pattern of different types of galaxies. As understood intuitively the astrophysics environment physics is intimated affected by cosmology priors with theoretical estimation or vise versa, or simply stating that the astrophysics effect couples the corresponding cosmology observations
   or the way backwards. In this article we try to quantify the influence on galaxy number density prediction at faint luminosity limit from the uncertainties in cosmology, and how much the uncertainties blur the detection of galaxy evolution, with the hope that this trying may indeed help for precise and physical cosmology study in near future or vise versa.
\end{abstract}

\maketitle

%========================================== INTRO ==========================================%
\section{Introduction}
Galaxy number count measures the number density $n$ of galaxies per unit solid angle $d\omega$ at redshift $z$ within the luminosity range $\left[L,L+dL\right]$ as~\cite{ls,rj}
\begin{equation}
n\left(\hat{r},L,z\right)dL~dz~d\omega = \phi\left(L,z\right)dL~d\hat{V} ,
\end{equation}
where $\phi(L,z)$ is the luminosity function (LF for short); $\hat{r}$ represents a specific direction on comoving coordinate with the corresponding differential volume $d\hat{V}$. The anisotropic metric is encoded in the direction, while for isotropic and homogeneous FRW metric, the relation between number density and LF is much simpler by integration over the solid angle and gives
\begin{equation}
n\left(L,z\right)~dL~dz = \phi(L,z) \frac{D_L^2}{(1+z)^5h^3 E}~dL~dz, \label{eq:n}
\end{equation}
where $E$, the dimensionless Hubble parameter at redshift $z$ equals to $H/(100 h~km\cdot s^{-1}\cdot Mpc^{-1})$ and the luminosity distance is defined as $d_L=\frac{c}{H_0}D_L$. The dimensional constant parameters are absorbed by normalization factor in luminosity function, leaving dimensionless terms. The luminosity functions are usually measured in the non-parametric way by astrophysicist, but that approach is based on pre-selected fiducial modeling of background cosmological evolution. It is believed that variations in cosmological parameters are too weak to be captured through the noisy observation, but we are driven by the optimism that sooner or later we are able to detect them.

%===================================== THEORETICALLY ======================================%
\section{Theoretical framework}
Luminosity function contains the intrinsic characteristic of different sets of galaxies, usually affected by redshifts, galaxy types and local environments. According to the assumption of the hierarchical structure formation with cold dark matter model, the most adopted analytical parameterization of luminosity function is the Schechter form~\cite{schechter} which reads
\begin{equation}
\phi(L)dL=\frac{\phi^\ast}{L^\ast}(\frac{L}{L^\ast})^\alpha exp(-\frac{L}{L^\ast})dL, \label{eq:lfL}
\end{equation}
with $L^\ast$, the characteristic luminosity and $\alpha$, the faint-end slope parameter. Other well-known luminosity function forms are
for example, the power-law model~\cite{lawrence}:
\begin{equation}
\phi(L)dL = \phi_\ast L^{1-\eta} (1+\frac{L}{\beta L_\ast})^{-\beta} dL,
\end{equation}
and log-gaussian model~\cite{saunders}:
\begin{equation}
\phi(L)dL = \phi_\ast (\frac{L}{L_\ast})^{1-\gamma} Exp[-\frac{\log_{10}^2(1+\frac{L}{L_\ast})}{2\sigma^2}]dL.
\end{equation}

When $L \ll L_\ast$, ``faint-end slope'' is a common feature of different luminosity functions. Even though there are other luminosity functions which have no slope at faint limit, or through non-parametric method where luminosity function is not assumed to be a specific form, the faint slope is still available as a parametric approximation. So we generalize the luminosity functions at faint luminosity limit as:
\begin{equation}
\phi(L)dL = \beta L^\alpha dL, \label{eq:general}
\end{equation}
with $\alpha$ and $\beta$ as constant parameters. The general form of luminosity function is convenient in discussing the cosmological effect as it will be shown in the following. With faint slope, the cosmological background effects in Eq.~(\ref{eq:n}) can be easily separated. At bright end, the number density can be significantly affected by gravitational lensing~\cite{er}, while background influence is suppressed by the rapid decrease of galaxy number.

From observation we capture the apparent magnitude of galaxies rather than absolute luminosity, where the cosmology is entangled with astrophysics. Absolute magnitude relates to the apparent one mainly through luminosity distance and K-correction as follows \cite{schmidt,eales,cminus,eep88,swml,rj,hogg,iribarrem,heyl}:
\begin{equation}
m = M + k + \mu(z),
\end{equation}
where $k$ represents the K-correction;  and the distance module $\mu(z)= 5\log_{10}(\frac{d_L}{10~pc}) =5 \log_{10} D_L - 5 \log_{10} h + 42.39$, in which $D_L$ is the Hubble free luminosity distance
that reads
\begin{equation}
D_L(z) = (1+z)\int_0^z\frac{dz'}{E(z')},
\end{equation}
where $E(z) = H(z)/ (100h km/s/Mpc)$. Eq.~(\ref{eq:general}) can then be expressed as
\begin{equation}
\phi(m,z)dm = \beta 10^{\alpha(k+\mu-m)}dm,
\end{equation}
by adopting $\frac{L}{L_\ast} = 10^{-0.4(M-M_\ast)}$. For simplicity, some constant parameters and values are absorbed in $\beta$ and $\alpha$ parameterizations.
Now the effect of cosmology can be separated from astrophysics, that is the luminosity function, which means
\begin{equation}
\phi(m,z) = (\beta 10^{-\alpha m}dm)\cdot 10^{\alpha(k+\mu)}.
\end{equation}
The number density then reads
\begin{equation}
n(m,z)dm dz = \phi(m)\frac{D_L^2 10^{\alpha(k+\mu)}}{(1+z)^5h^3E}dm dz, \label{eq:final}
\end{equation}
where $\phi(m) = \beta 10^{-\alpha m}$. Eq.~(\ref{eq:final}) can be integrated analytically, for predicting
galaxy number within given range of apparent magnitude and redshift. We emphasize that this method must be
limited within the faint-end of galaxy survey samples. The evolution of galaxy is also coupled with cosmological
model normally, which will bring large uncertainty in estimating cosmological parameters through galaxy number count.

%================================== NUMERICALLY ========================================%
\section{Numerical analysis}
In order to estimate the cosmological influence on galaxy number count at faint limit, we compare the consequent error in number density prediction of galaxy generated from variations in cosmological parameters, with respect to the fiducial $\Lambda CDM$ model where $H^2(z) = H_0^2[\Omega_{m0}(1+z)^3+1-\Omega_{m0}]$. We define the comparison ratio $\gamma_n$ which reads
\begin{equation}
\gamma_n = \frac{n}{n_{(fid)}} = \gamma_{D_{L}}^{(5\alpha +2)} \gamma_h^{-(5\alpha +3)} \gamma_E^{-1}, \label{gamma}
\end{equation}
where $n$ indicates density prediction within the possible distribution range of cosmological parameters, and $\gamma_{D_{L}} = D_L/D_{L(fid)}$, $\gamma_h = h/h_{(fid)}$ and $\gamma_E = E/E_{(fid)}$ as well. The elements with subscript ``(fid)'' are calculated according to fiducial model.  The parameters of luminosity function $\phi(m)$ are set as the same since only cosmological effect is considered here, but inevitably, the ``faint-end-slope'' $\alpha$ can not be removed, which has impact on the result. However, we roughly know about the range where the value of the slope lies from previous researches with samples observed at low redshift ($0.05<z<0.2$) and intermediate redshift ($z<1$)~\cite{sapm,cfa,sdss1,sdss6,bell,zucca,methods}.
In the following, we will display $\gamma_n$ in different cases.

In the first case, we adopt the $\Lambda CDM$ model and the variations come from the Planck+WP+highL result~\cite{planck13}. The fiducial model is set with best-fit values of parameters, while the uncertainty in the estimations of parameters is taken to produce $\gamma_n$. Following this setting, we have
\begin{eqnarray}
\gamma_{D_{L}} &=& 1+\delta_{\Omega_{m0}}(\frac{\partial D_L}{\partial \Omega_{m0}})/D_{L(fid)},\\
\gamma_E &=& 1+\delta_{\Omega_{m0}} \frac{(1+z)^3 -1}{E^2_{(fid)}},\\
\gamma_h &=& 1+\delta_h/h_{(fid)},
\end{eqnarray}
where by taking the $1\sigma$ limits of the variations of cosmological parameters, $\Omega_{m0} = 0.315\pm 0.017$ and $h = 0.673\pm 0.012$, which means  $ that\delta_{\Omega_{m0}} =\pm 0.034$ and $\delta_h =\pm 0.024$.

\begin{figure}\label{fig1}
\begin{center}
\includegraphics[width=0.35\textwidth]{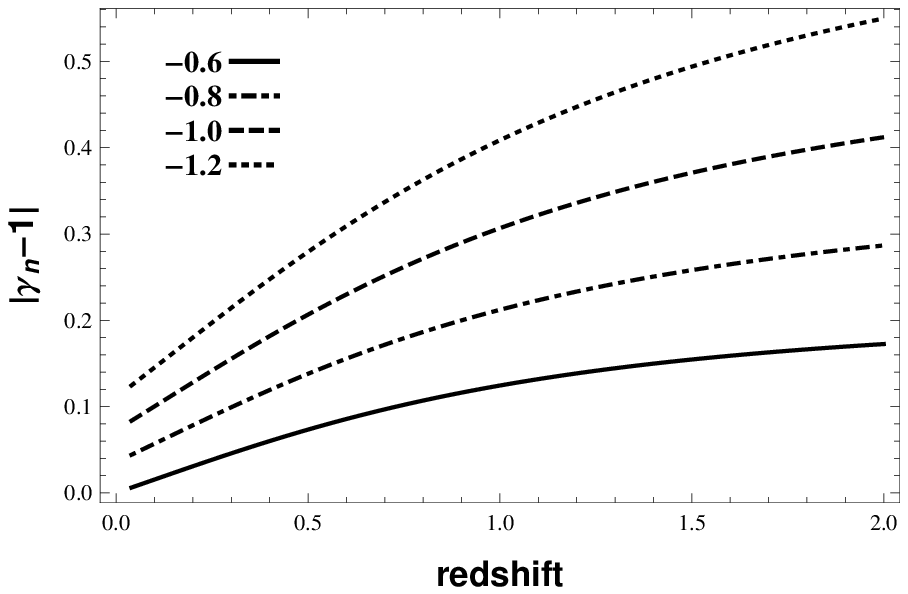}\\
\includegraphics[width=0.35\textwidth]{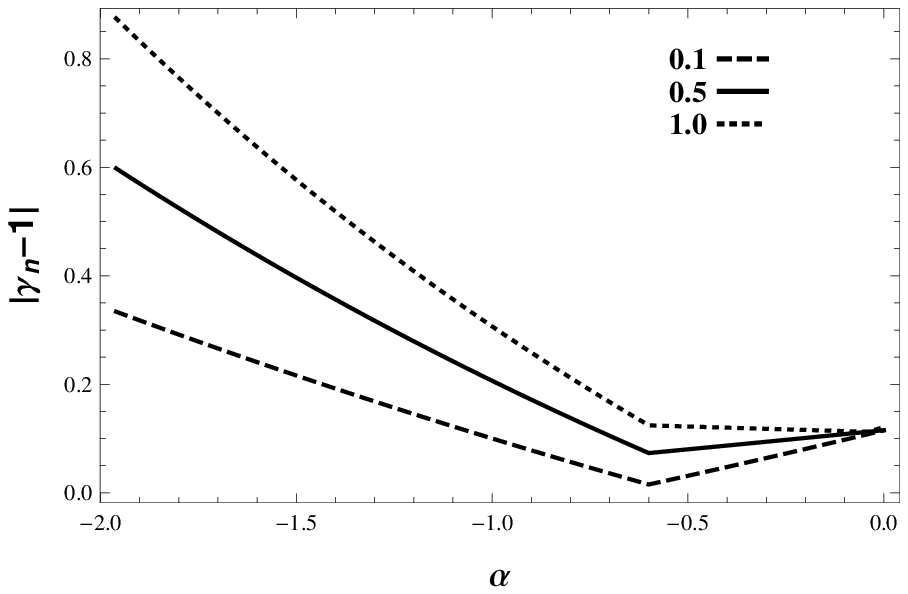}
\end{center}
\caption{Comparison ratio $\gamma_n$ derived according to uncertainties in cosmological parameters of $\Lambda CDM$ model, the
upper panel shows redshift dependency of estimation error with different faint-end slopes, while curves in the lower panel represent variation according to $\alpha$ at fixed redshifts.}
\end{figure}

As we can observe in Fig.1, estimation error of galaxy number density generally goes up with higher redshift and steeper faint-end slope, brought by the uncertainty of cosmology estimation with standard $\Lambda CDM$ model according to CMB observations.

Recently, some research papers suggest the possible deviation from standard $\Lambda$CDM model~\cite{gbz,salv}, which mainly comes from comparison and combination of low-redshift and intermediate-redshift observations. Union$2.1$ data analysis~\cite{union21} suggests that $\Omega_{m0} = 0.278^{+0.020}_{-0.019}$ at $1\sigma$ limit for $\Lambda$CDM model, and $h=0.738\pm 0.024$ was estimated in ref.~\cite{hst}. The result in parameter-frame $\{\Omega_{m0},h\}$ given by SNe Ia observations is truly inconsistent with those estimated according to CMB observations at $2\sigma$ level. Notice that physically meaningful parameter-set should be $\{\Omega_{m0}h^2, h \}$, where $\Omega_{m0}h^2$ is estimated to be $0.14187\pm 0.00287$ from Planck's report, while to be $0.137\pm 0.007$ from Union$2.1$ with $h = 0.702\pm 0.026$. Interestingly, the inconsistency is relieved to be     $1\sigma$  level after changing the parameter-frame in the same parameter space. This is reasonable, as $\Omega_{m0}h^2$ represents the quantity of physical dust matter density while $h^2$ in the same frame represents physical density of total energy in the flat base $\Lambda$CDM model; meanwhile, in the $\{\Omega_{m0},h\}$ frame, $\Omega_{m0}$ indicates the contrast of matter density against total energy density, which means $\Omega_{m0}$ is much more sensitive to matter-to-field transition than $\Omega_{m0}h^2$.

%===================================== UNDER CONSTRUCTION ============================================%

Without convincing conclusion about cosmology deviation from standard $\Lambda$CDM model \cite{wm,pal}, we can also address the coupling between non-standard cosmology and galaxy density evolution, which reads
\begin{equation}
(1+z)^\epsilon = \frac{(\tilde{D_L}/D_L)^210^{\alpha (\tilde{\mu}-\mu)}}{(\tilde{E}/E)},
\end{equation}
where $\epsilon$ represents possible fraction of galaxy density evolution (assumed to be a power-law form) caused by cosmological influence, the cosmological quantity with tilde overhead comes from prediction with non-standard model, while the ones without represent standard predictions. Notice that the right-hand-side of the above equation is exactly what we had defined in Eq.~(\ref{gamma}), so the coupling is about power-law parameterization of $ the \gamma_n$, which reads then

\begin{equation}
\gamma_n = (1+z)^\epsilon .
\end{equation}

%======================================== CONCLUSION ================================================%
\section{Conclusion}
In this short letter, we briefly observed the influence from cosmology on predicting galaxy number densities. The observational uncertainty in $\Lambda$CDM model has negligible effect at low redshift but the uncertainty may raise over 5 percent at redshift
higher than $0.5$. We also evaluated the effect in a model independent way, with low redshift observational datasets. The relative difference against fiducial $\Lambda$CDM model is about $10$ to $20$ percent. In current work, we have not included the coupling issue between cosmology and galaxy evolution into discussion, which lies beyond this article. From the latest Planck results \cite{pal} the $H_0$ value and fluctuation amplitude tensions still exist, which may imply we need new physics beyond the base $\Lambda$CDM model if the current results hold up and  further observations enhance so. We hope the quantifying of cosmological influence on galaxy number count based on current observation is helpful in more robust estimation of galaxy evolution and the corresponding cosmology study in the near future. With the upcoming accuracy datasets besides the Planck's, SKA's and SDSS's we optimistically believe with sufficient confidence that
 this kind of work is full of positive effects on the precise cosmology research  with very promising future .

\section*{Acknowledgments}
We thank devoted J X Wang for great help for this long time discussed work. Fruitful discussions on related topics of XHM with Prof. R Sheth during ICTP cosmology activities are highly appreciated.
This work is partially supported by the Natural
Science Foundation of China (NSFC) under Grant No.11075078.

\end{document}